# Comprehensive process-molten pool relations modeling using CNN for wire-feed laser additive manufacturing

**Authors**

Noopur Jamnikar[a,b], Sen Liu[a,b], Craig Brice[a,b], and Xiaoli Zhang[a,b*]
[*]Corresponding authors: xlzhang@mines.edu

**Affiliations**

[a] Mechanical Engineering, Colorado School of Mines, Golden, CO 80401 USA
[b] The Alliance for the Development of Additive Processing Technologies, Colorado School of Mines, Golden, CO 80401 USA

**ABSTRACT:** Wire-feed laser additive manufacturing (WLAM) is gaining wide interest due to its high level of automation, high deposition rates, and good quality of printed parts. In-process monitoring and feedback controls that would reduce the uncertainty in the quality of the material are in the early stages of development. Machine learning promises the ability to accelerate the adoption of new processes and property design in additive manufacturing by making process-structure-property connections between process setting inputs and material quality outcomes. The molten pool dimensional information and temperature are the indicators for achieving the high quality of the build, which can be directly controlled by processing parameters. Further, the molten pool dimensional and thermal profiles result in a printed bead with various geometric shapes and microstructural characteristics. For the purpose of in situ quality control, the process parameters should be controlled in real-time based on sensed information from the process, in particular the molten pool. Thus, the molten pool-process relations are of preliminary importance. This paper analyzes experimentally collected in situ sensing data from the molten pool under a set of controlled process parameters in a WLAM system. The variations in the steady-state and transient state of the molten pool are presented with respect to the change of independent process parameters. A multi-modality convolutional neural network (CNN) architecture is proposed for predicting the control parameter directly from the measurable molten pool sensor data for achieving desired geometric and microstructural properties. Dropout and regularization are applied to the CNN architecture to avoid the problem of overfitting. The results highlighted that the multi-modal CNN, which receives temperature profile as an external feature to the features extracted from the image data, has improved prediction performance compared to the image-based uni-modality CNN approach.

## 1. Introduction and motivation

Additive manufacturing (AM) is a trending topic in the current manufacturing industry due to the high level of automation, real-time performance monitoring and control, and high yield of production capability [1]. Wire feed laser additive manufacturing (WLAM) is in the family of directed energy deposition (DED) AM processes, in which a focused energy source and feedstock material are concentrated at a focal point, commonly in the presence of inert gas. In DED AM process, energy is guided to a narrowly focused area, mainly powered by a source such as a laser beam, electron beam, plasma arc, etc., to melt the deposited materials along with the substrate [2, 3]. The material generally used is either in powder or wire form. The material is deposited in a layer by layer form until the part reaches a near-net shape. In this paper, we focus on the laser power source concentrated on the substrate to melt the heated wire from the wire feeder in the presence of argon gas. There are many interrelated parameters involved in the process, and their effect on the molten pool characteristics, droplet kinematics, and thermal history is significant for determining surface roughness, texture, and part geometry [4, 5]. The high complexity of AM process and associated cost represents the main barriers to the widespread adoption of WLAM as a mainstream manufacturing method. To solve these issues, advanced technologies for sensing, modeling, and control are needed to improve the accuracy and process stability of WLAM.

The WLAM process involves complicated interactions between laser power, wire supply, substrate, and local inert gas environment. There has been a continuing effort to develop analytical, numerical, and data-driven models for describing the molten pool-process relations during AM processes [6-13]. Traditionally, process parameters such as laser power and scan speed are held constant with time. The molten pool temperature will increase as the process continues due to accumulated residual heat from previous layers. Without process feedback control, non-uniform build morphology will form, such as increasing bead width, uneven bead height, increased molten pool depth and dilution, and thermal distortion due to the build-up of residual stress [4, 14]. Also, fluctuations in processing parameters (e.g., laser power, scan speed, and material supply rate) often force the process parameters to deviate from the pre-defined condition, resulting in defects in the printed parts [15, 16]. Therefore, establishing relationships between the molten pool and process parameters is necessary for a real-time monitoring and feedback control system that can improve the quality of AM fabricated parts.

The existing studies for monitoring and controlling AM systems focus on molten pool geometry, thermal profile, and deposition control. Different measuring systems, including a photodiode, pyrometer, and camera, were used to detect the temperature radiation. Sensing and controlling molten pool size over a range of process parameters is critical for maintaining consistent bead shape and ensuring geometrical accuracy of the produced part [17, 18]. Charged coupled device (CCD) cameras have been extensively used to obtain molten pool dimensional information [15, 19, 20]. Process maps have been developed for predicting the steady-state molten pool size at a step-change in laser power or laser velocity with numerical models [21, 22]. The process map for controlling solidification microstructure is achieved through direct molten pool dimension control, with significant implications for AM processes [23, 24]. The key sensors used to monitor and control the molten pool thermal profiles are pyrometers and thermocouples, which provide a wide temperature measurement range and non-contact measurement capability [25-27]. The infrared pyrometers coupled with digital cameras are the primary and desirable option for monitoring and controlling the molten pool during AM. For example, Hua et al. [28] used a two-color infrared thermometer to investigate the influence of laser power on the molten pool temperature during the laser AM process. Wang et al. [6] built a physics-based multivariable modeling and feedback controller to predict molten pool geometry and temperature. Tang et al. [14] designed a molten pool temperature controller to achieve a constant or time-varying temperature even when the process parameters vary significantly. Song et al. [16] developed a two-input single-output hybrid control system to control both height growth and molten pool temperature at each deposition layer. Hofman et al. [29] controlled the molten pool width using laser power through the camera's feedback data for a laser cladding system. Suh et al. [30] developed a proportional-integral-derivative (PID) and fuzzy logic-based model to control the height of the deposit using an in-process monitoring camera that commanded changes in laser power based on measured information from the camera.

With the recent advancements in deep learning, convolutional neural networks (CNN) have emerged as a robust method in the field of computer vision, such as pattern recognition, classification, and detection algorithms. CNN converts the features from the input image into an abstract representation for regression and classification tasks. CNN and their variations have recently been used in the AM industry for classification [31-33] and regression [34-39] problems. For example, Gonzalez et al. [35] presented a novel convolutional laser-based manufacturing (ConvLBM) method to extract features from images for a laser-based manufacturing process in real-time to predict laser power and laser scan speed. Kwon et al. [37] used a CNN for predicting the laser power of the powder bed fusion process using molten pool images. Knaak et al. [38] presented a CNN model to segment the molten pool images and detect the deviation in the current process parameter, i.e., laser power and welding speed, using supervised machine learning algorithms.

Despite these fundamental and practical applications for in situ quality assurance and control, the influence of process parameters on the dimensions and thermal behavior of the molten pool has not been investigated systematically [28]. Most of the above work done in the field of AM process is limited to modeling the correlation of single sensing modality and single process parameter. This is due to modeling technique limitations for handling the inter-dependent correlation complexity and the high cost of experiments. Single sensing modality-single process parameter modeling is relatively simple as it reduces the number of control processing variables and provides limited capabilities for comprehensive quality control [30, 40-43].

The limitations of single sensing solutions motivates the need to include more process parameters in the architecture to achieve a broader range of operation capability and robust control. WLAM is a complicated process, and developing a single modality-single process model may not capture the inter-dependable relationship between the process and sensing parameters. Monitoring the top surface molten geometry does not provide sufficient information for the molten pool depth and thermal profiles [4, 44]. The deposited bead geometry is directly correlated to molten pool dimensions, while the as-deposited microstructure is correlated with the molten pool thermal profiles such as leading and trailing edge temperature and re-melt ratio (the relative amount of previously deposited material that is remelted during subsequent layers). More comprehensive sensing of the molten pool that fully captures the physical effects of multiple controllable process parameters is needed to improve the quality of AM fabricated part, both in terms of geometry and performance.

To meet the challenge of in situ quality control during the WLAM process, we proposed a multi-modal multi-output CNN approach to explore the relation between in situ process sensor data and controllable process parameters. The network takes in molten pool image and temperature profile data as input to predict the corresponding process parameters. This end-to-end prediction model eliminates the need for off-line feature extraction and selection processes, which directly takes sensor data for the purpose of in situ quality assurance. The results are highlighted as follows: 1) Experimental data is collected from a single layer printed under a set of controlled process parametric combinations. Molten pool image and temperature information are measured using a CMOS camera and multiple pyrometer sensors. 2) The effects of process parameters on molten pool features and their correlations are calculated. The molten pool-process relation is formulated as a data-driven multi-modal CNN regression model. 3) Sensitivity analysis is performed to study the effect of change in the independent process parameters on the sensing data for monitoring and control decisions. 4) It is proven that the additional temperature feature as input along with the image feature improves the prediction performance of the molten pool-process relations model. The performance of the developed end-to-end CNN architecture is evaluated by measuring the convergence and the prediction performance with different input feature combinations.

## 2. Materials and methods

### 2.1. WLAM experimental setup

The WLAM DED system has been developed and installed at Oak Ridge National Laboratory in Knoxville, Tennessee. A 6kW laser is delivered to the end effector of the robot arm in the presence of argon filled environment. The feedstock is 1.5875 mm Ti-6Al-4V welding wire per AMS 4954K specification. The laser WLAM DED robot setup, along with the mounted sensors, is shown in Figure 1. The sensors were selected to capture as much data as possible from the process during operation. There were five categories of data that were collected: (1) visual, (2) thermal, (3) positional, (4) chemical, and (5) acoustic. Two Prosilica GT1930C cameras were mounted to the robot head, one directly coaxial with the process and the other at a 90° oblique angle to the primary direction of travel. The camera is connected using an Ethernet interface, recording $1936 \times 1216$ pixels images at 25 frames per second (fps) using the NI PXIe-8234 vision module. The main issue in the laser-based AM process is that the image contrast for the molten pool is too bright to capture the surface morphology directly. Hence, bandpass filters are mounted in front of the camera to reduce the intensity. There are three pyrometers with a temperature range of 50-400°C, 200-1500°C, and 1000-2000°C to measure the leading (Optris CTlaser 3M), trailing (Optris CT XL 3M), and molten pool (Optris CTlaser 05M) temperatures, respectively. The pyrometers are calibrated for emissivity using a heated plate and physical contact measurements with thermocouples. Note that the molten pool pyrometer could not be easily calibrated using a similar method, so the presented data are considered relative and not absolute. The leading and trailing edge pyrometers are pointed approximately 25 mm in front of and behind the molten pool. Temperature data is collected at 100 Hz using NI PXIe-4302 analog input module. An acoustic sensor is mounted on the laser head operating a frequency of 1 kHz to capture variations during the build. Analysis of the sensor signals relative to the process during stable and unstable operation will determine how each sensor can be used in the control logic. The National instrument (NI) industrial controller NI PXIe-8880 along with the vision development module and analog/digital I/O module, is used for monitoring and controlling the laser DED system. This paper focuses on studying the in situ sensing data from the coaxial camera and three pyrometers.

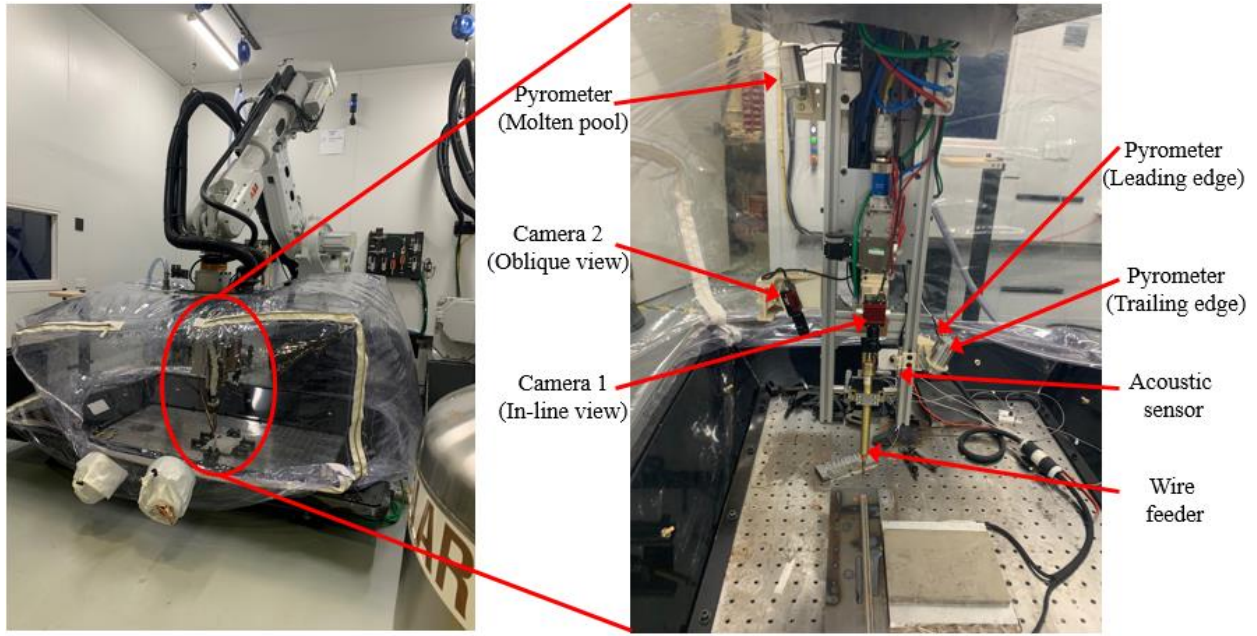

Figure 1. Integrated laser hot wire feed DED system.

## 2.2. Data preparation

In this study, thirteen 100 mm single bead deposition experiments were conducted for different values of process parameters; laser power, travel speed, wire feed rate, and hot wire power. Table 1 shows the combination of process parameters used for data collection for the WLAM system. The process parameters vary in the range of 4000-6000 W, 3.5-10 mm/s, and 40-71.3 mm/s for laser power (LP), travel speed (TS), and wire feed rate (WFR), respectively. The single bead deposition data consists of both the steady and transient state of the molten pool. In this paper, we will focus on the steady-state molten pool data. The pixel range for the image data from the coaxial camera is enormous, and it covers a vast portion of the unwanted region. The molten pool image is preprocessed with a selected region of interest (ROI) and cropped to reduce the image size without missing any information and keeping the data within the hardware processing capability. Hence, the coaxial camera images are cropped to a size of 481× 566 to be processed and trained in MATLAB. The dataset consists of 6500 images from 13 builds, containing 500 stable molten pool images from a coaxial camera and temperature data from each build. From the 13 build dataset, 12 builds are used for training the network, while the remaining unseen one build is used in the testing phase.

Table 1. Process parameter setting for the collected image and temperature data.

| Experiment # | Process parameters | | | |
| --- | --- | --- | --- | --- |
| | Laser power (W) | Travel speed (mm/s) | Wire feed rate (mm/s) | Hot wire power (W) |
| 1 | 6000 | 6.6 | 60 | 300 |
| 2 | 6000 | 5 | 60.1 | 300 |
| 3 | 6000 | 3.5 | 71.3 | 400 |

| 4 | 6000 | 3.5 | 50.1 | 200 |
|---|---|---|---|---|
| 5 | 4500 | 5 | 48.4 | 300 |
| 6 | 4500 | 5 | 50.8 | 300 |
| 7 | 4500 | 10 | 50.8 | 300 |
| 8 | 4500 | 5 | 43 | 300 |
| 9 | 4500 | 5 | 40 | 300 |
| 10 | 5000 | 5 | 40 | 300 |
| 11 | 4000 | 5 | 40 | 300 |
| 12 | 4500 | 6.5 | 40 | 300 |
| 13 | 4500 | 3.5 | 40 | 300 |

*2.3. Feature correlation analysis*

By increasing the number of features, feature engineering increases the problem's dimensionality, leading to the "curse of dimensionality" [45, 46]. Hence, before final tuning and verification of the machine learning models, it is desirable to analyze redundant or insignificant inputs. In such a case, the number of inputs may be reduced, reducing the functional complexity of the machine learning models. Feature selection is an essential technique for improving the network's performance by finding the most meaningful input features for the output prediction. It helps in improving the accuracy caused by the redundant and enormous amount of available experimental data. It is not necessary that the addition of more features into the network always lead to a better outcome. The real-world environmental data could be noisy, redundant, and invalid or have disturbances caused by the process or environment. The data used for building the model might not always be able to recognize any real pattern between the mappings. Furthermore, the corrupted data could increase network complexity in terms of time and storage. Hence, it is necessary to analyze feature pairs correlation before training the CNN model. Features correlation and redundancy were evaluated using Pearson correlation [47]. The Scikit-learn python implementation of these algorithms was used [48]. Correspondingly, Pearson correlation between feature pairs $r_{x_{ij}}$ or feature and property $r_{xy}$ uses the standard definition,

$$r_{xy} = \frac{\sum_{i=1}^{n}(x_i-\bar{x})(y_i-\bar{y})}{\sqrt{\sum_{i=1}^{n}(x_i-\bar{x})^2}\sqrt{\sum_{i=1}^{n}(y_i-\bar{y})^2}} \tag{1}$$

Where, $n$ is the sample size, $x_i$ and $y_i$ are the individual sample points, and $\bar{x}$ and $\bar{y}$ are the sample means.

*2.4. Sensing-process modeling*

Figure 2 shows the sensing-process (S-P) relations model, employing a convolutional neural network that takes in the sensing data from the CMOS camera and pyrometer as inputs and outputs the process parameters to achieve it. Firstly, data collection was conducted on a WLAM system installed with sensors to record the molten pool evolution during printing. The effects of process parameters on molten pool features and their correlations are visualized and analyzed. Next, the sensing image data collected during the build are paired with corresponding molten pool temperature data points as shown in the input block. Third, CNN architecture automatically performs feature extraction without human intervention via convolution and pooling layers to obtain meaningful features from the molten pool images. Finally, the fully connected layer takes input extracted features in conjunction with the molten pool temperature for final process parameter estimation, i.e., the laser power, travel speed, and wire feed rate. The network is trained on the data collected on a WLAM system under a set of controlled process parameters. Once the network is trained, it can be used to predict the process parameters given unseen sensor data.

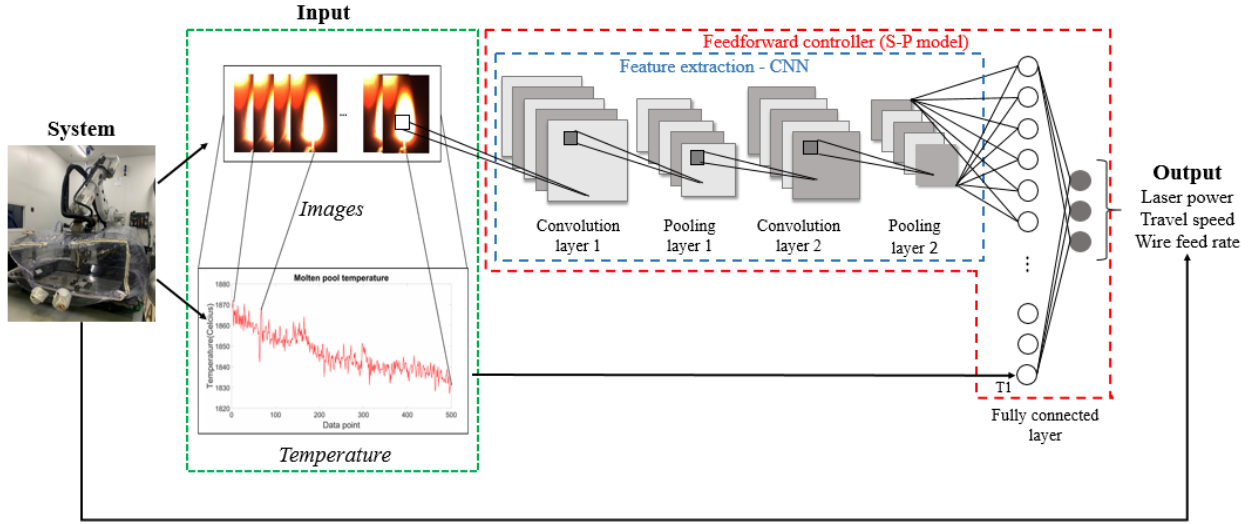

Figure 2. Sensing-process modeling using CNN for control parameter prediction using molten pool images and temperature data.

Specifically, a convolutional neural network is based on an artificial neural network, a popular machine learning method inspired by neurons in the human body. In recent years, a plethora of work has been done using CNN on experimental data for various applications [49-52]. CNN is widely used in image processing because it learns spatial local relations by exploiting the features in a hierarchical nature. CNN takes advantage of the multi-dimensional image input, shares weights along with the layers, and knows the intricate features concerning each other from high-dimensional image data to output classes or values. CNN consists of different layers such as a convolutional layer, activation function, pooling layer, fully connected layer, batch normalization, dropout, etc., for classification and regression tasks. The input image is first processed by a set of kernels, also known as filters. The convolutional layer is applied to extract features by convolving the features from the previous layer and shifting the filter as specified by the stride. The filters are learnable weights that are continuously updated during learning, while the size of the filter is user-defined. The feature map F is represented in Eq. (2) as,

$$F(i,j,k) = \sum_{m=0}^{M-1}\sum_{n=0}^{N-1}\sum_{l=0}^{L-1} E(i+m, j+n, l) K_k(m.n,l) + b_k \qquad (2)$$

Where, $E$ is the extracted features from the previous layer, $K_k$ is the applied filter, $b_k$ is the bias, (M, N, L) is the size of the filter, and L is the 3rd dimension of the previous feature map layer.

The feature map represents the input, i.e., the molten pool image response to the feature defined by the input filter. With the addition of more convolutional layers, the features become more abstract and directed towards the output prediction/classification problem. After the convolution layer, a non-linearity activation function is applied to the network to introduce non-linearity. The standard activation functions that are applied are ReLU, Tanh, Sigmoid, LeakyReLU, etc. In this work, the hyperbolic tangent (tanh) activation function is used. A batch normalization layer is applied to address the issue of covariant shift and non-uniform scaling ranges in the internal layers. This regularization layer reduces overfitting and network malfunction due to improper learning rate and speeding up the training process [53]. Pooling layers are added to downsample the data dimension and introduce local invariance, making the feature map less sensitive to locations. Standard pooling functions are global pooling, average pooling, max pooling, etc. The pooling layer decreases the computational volume of data and increases the robustness of the algorithm. The last pooling layer's output is an input feature vector to a fully connected layer to predict the final process parameters. The output of the fully connected layer is given by Eq. (3) as,

$$X_{i+1} = W_i . X_i + b_i \qquad (3)$$

Where, $X_i$ is the i$^{th}$ fully connected layer, $W_i$ is the weight matrix, and $b_i$ is the bias vector.

The different layers are connected in a user-defined manner to form a network that is trained using backward propagation that uses a gradient descent technique. The learning parameters are adjusted during the training to achieve accurate prediction of process parameters and minimize the training loss function, i.e., mean square error in our case as given in Eq. (4),

$$Loss_{MSE} = \sum_1^N l(y^n, \hat{y}^n)/N \qquad (4)$$

Where, $l$ is the loss function, $y^n$ is the target value, $\hat{y}^n$ is the predicted value, and N is the size of the training data.

*2.5. Model Evaluation*

In order to predict the process parameters accurately, the network structure, input, and output need to be evaluated along with the structure generalization capability. The generalization is gauged by performing 6-fold cross-validation where is the network is trained on 6000 molten pool image and temperature data and tested on the 500 unseen samples. When constructing a CNN, the more input layers result in a higher quality of the extracted features. However, this is not always the case because of dissipating or shooting gradient or degenerative models when using deeper networks. When training a deep NN, the gradient calculated using backpropagation is responsible for updating the network's weights. Poorly designed network structure causes the gradient to shoot or vanish, resulting in a slower model learning process and lower accuracy. Hence, the CNN architecture and the layers are optimized to suit the specific application for designed input and output. We optimized the CNN for two variations of input type, which includes just the molten pool image as input, i.e., uni-modality and molten pool image, and temperature as inputs, i.e., multi-modality.

The evaluation criteria used for performance comparison are the root mean square error (RMSE), standard deviation (SD) for RMSE, and relative percentage error (RE) between the actual and predicted parameters, given in Eqs. (5, 6) respectively. $y_i$ is the measured, and $y_i'$ is the model predicted process parameter. Normalized RMSE (NRME) is another error evaluation metric for different scales of process parameters, as given in Eq. (7). The RMSE is normalized to 2000 W laser power, 6.6 mm/s travel speed, and 31.3 mm/s wire feed rate.

$$RMSE = \sqrt{\frac{1}{n}\sum_{i=1}^{n}(y_i' - y_i)^2} \qquad (5)$$

$$RE = \frac{1}{n}\sum_{i=1}^{n}\left|\frac{y_i' - y_i}{y_i}\right| \times 100\% \qquad (6)$$

$$NRMSE = \frac{RMSE}{(y_{max} - y_{min})} \times 100\% \qquad (7)$$

## 3. Results

### 3.1. Sensing-process relations description

The molten pool is a region of superheated molten metal in proximity to the laser/material interface-typically in the form of a hemi-spherically shaped droplet and moves at the robot travel speed. Since the molten pool determines the outcome microstructure and properties, its morphology, temperature, and wetting behavior are of paramount interest in quality control. In this part, a molten pool sensing WLAM system has been developed by using three infrared pyrometers and one CMOS camera. Real-time tracking and measurement are achieved during the printing processes. The influence of laser power, travel speed, and wire feed rate on the molten pool characteristics are investigated in Figure (3-5). As described in section 2.3, only the steady-state molten pool sensing data from the 13 build samples are used to train and test the network. The molten pool images are paired with corresponding temperature profile data from the pyrometers to be used for sensing-process relations modeling.

#### 3.1.1. Effect of change in laser power

Figure 3 presents the effects of laser power on the molten pool's characteristics at a fixed robot travel speed of 5 mm/s and wire feed rate of 40 mm/s. Figure 3(a-c) shows the typical molten pool size variations, and Figure 3(d) shows the molten pool temperature at a steady printing state for 0-200 seconds internal. Note there is a reflection of the molten pool due to the lens filters in the upper left of each image; this reflection is ignored during image processing. The molten pool size increases with laser power from 4000 to 5000W, as expected due to the enlarged heat-affected zone at higher laser energy density. The molten pool temperature holds steady at a temperature of 1800°C at LP = 4000 W. As laser power increases to 4500 and 5000 W, the corresponding temperature increased to 1850°C and 1925°C, respectively. The suppled Ti-6Al-4V wire is fully melted (liquidus temperature of 1660°C). It can also be observed that there is a slight decreasing trend of temperature for higher laser power. It may result from the experimental setup of sensors and data collection uncertainty caused by the deviation of the sensor's focal point from the center of the molten pool progressively during the printing process. It has been previously documented in studies directly measuring molten pool temperature that the temperature in the molten pool can vary widely depending on the radial distance from the beam impingement point [54].

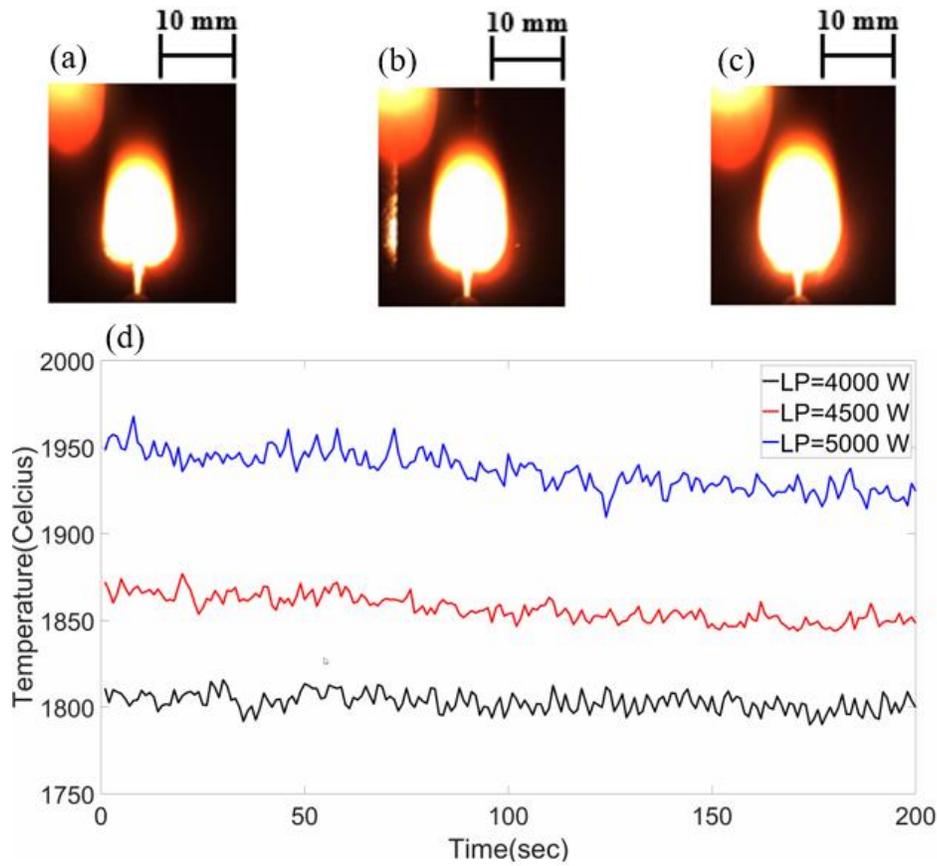

Figure 3. Illustration of samples for stable molten pool image and the corresponding molten pool temperature data for change in laser power at a fixed travel speed of 5 mm/s and wire feed rate of 40 mm/s. (a) LP = 4000 W, (b) LP = 4500 W, (c) LP = 5000 W, (d) Molten pool temperature profile.

*3.1.2. Effect of change in travel speed*

Figure 4 illustrates the molten pool variation with the change in travel speed measured at a fixed laser power of 4500 W and wire feed rate of 40 mm/s. Figure 4(a-c) shows that the typical molten pool length slightly decreases with increased travel speed from 3.5 to 6.5 mm/s. This is due to a decrease in time-dependent energy density with the increased travel speed. While the laser power had a strong influence on the molten pool temperature, the travel speed shows a minor influence on the temperature. At travel speed 3.5 and 5 mm/s, the temperature signals are almost the same in the range of 1840-1870°C. At a speed of 6.5 mm/s, it can be seen that the temperature signal increases by about 20°C in the entire time series. This conflicts with the intuition that the molten pool temperature should decrease with the increase in laser travel speed. The underlying reason should be the range of travel speed variation, which is limited to cause a significant change in the laser energy input. The uncertainty in the experimental setup could be an additional contributing factor. The molten pool temperature has lower dependence on the travel speed than the laser power, as reported in the literature [25, 26, 55]. Bi et al. [25] observed that the molten pool temperature measured by a quotient pyrometer increases a little from travel speed 5 to 8.33 mm/s, then a slight decrement is observed up to a speed of 16.67 mm/s.

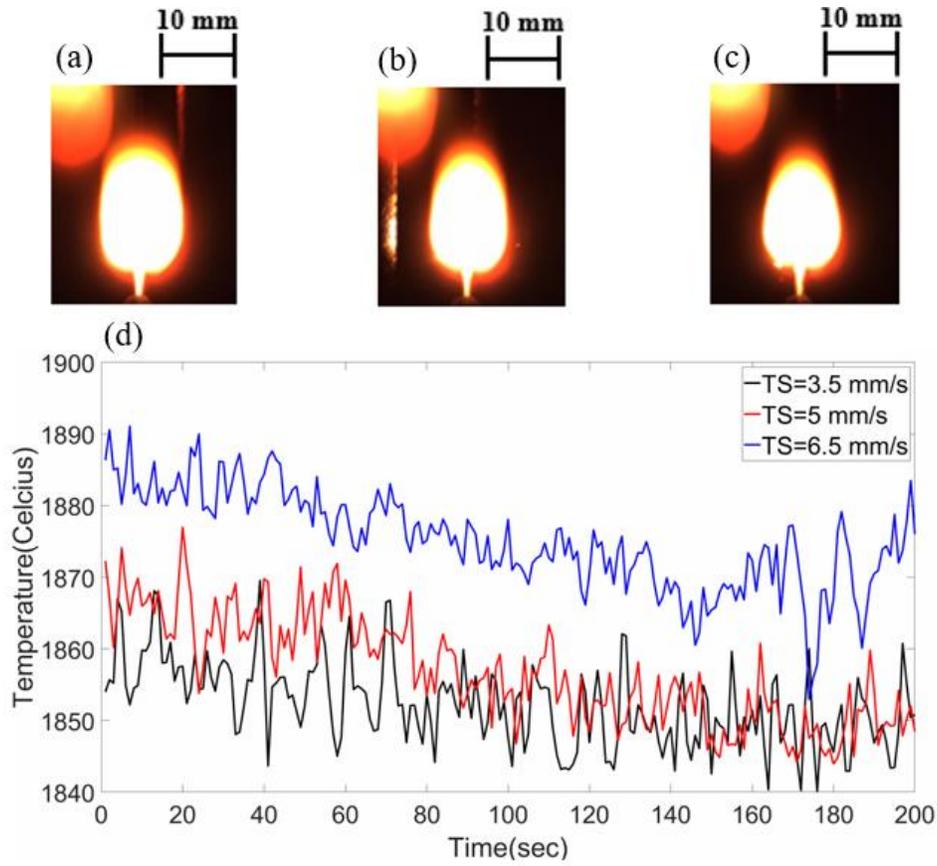

Figure 4. Illustration of samples for stable molten pool image and the corresponding molten pool temperature data for change in travel speed at a fixed laser power of 4500 W and wire feed rate of 40 mm/s. (a) TS = 3.5 mm/s, (b) TS = 5 mm/s, (c) TS = 6.5 mm/s, (d) Molten pool temperature profile.

*3.1.3. Effect of change in wire feed rate*

Figure 5 shows the effects of wire feed rate on characteristics molten pool at a constant laser power of 4500 W and travel speed of 5 mm/s. The dependence of molten pool features on the wire feed rate is complex compared to laser power and travel speed because of non-linear physical effects. When the laser power and wire pre-heat power are high enough, the increase of wire feed rate will enlarge molten pool size and increase molten pool temperature. Whereas with insufficient laser energy input, the molten pool size and temperature will drop, resulting from the cooling effect of the solid feedstock wire entering the molten pool. Figure 5 shows the molten pool temperature dropped with the increase of the wire feed rate. It results from the cooling effect of the solid feedstock, and the temperature of the feedstock from the wire preheater has an additional effect. Correspondingly, it can be observed that the molten pool surface profile stays almost constant at 40 and 48.4 mm/s, then a smaller molten is observed due to the lower molten pool temperature.

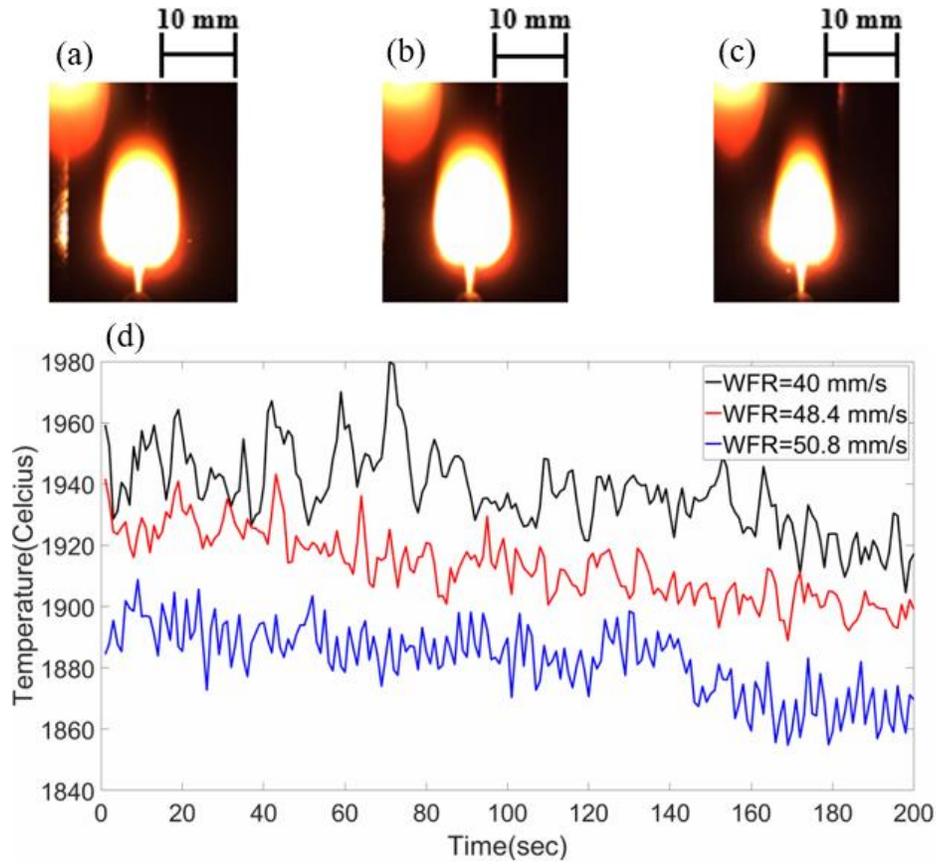

Figure 5. Illustration of samples for stable molten pool image and the corresponding molten pool temperature data for change in wire feed rate at a fixed laser power of 4500 W and travel speed of 5 mm/s. (a) WFR = 40 mm/s, (b) WFR = 48.4 mm/s, (c) WFR = 50.8 mm/s, (d) Molten pool temperature profile.

*3.2. Sensitivity analysis*

Sensitivity analysis (SA) is an integral component of any modeling, monitoring, and control system since it is used to investigate the complex relationship between different parameters systematically. The analysis involves all the four controllable parameters, i.e., laser power, travel speed, wire feed rate, hot wire power (HWP), and their corresponding outcome on the measurable sensing data. The effect of change in the controllable parameter is analyzed for significant change in the molten pool width and temperature data. It allows us to identify the smallest change in input that will provide a considerable change in output detected by the system. It also helps in identifying the measure of the error in terms of prediction accuracy evaluation. The smallest change in the controllable parameter that affects the measurable/response parameter is used as a measure bound for performance prediction in terms of model viability and precision. This study uses the operational acceptance testing (OAT) approach, where only one variable is changed from the nominal value and returned back. The nominal set values of the process parameter are laser power = 4500W, travel speed = 10mm/sec, wire feed rate = 40mm/sec, and hot wire power = 300W. The sensitivity study involved changing the laser power to a value of ± 225 W and ± 500 W from the nominal value while keeping the remaining parameters constant. The same technique was used for changing the TS, WFR, and HWP in a step of ± 0.5 mm/s and ± 1.5 mm/s, ± 4 mm/s and ± 10 mm/s, and ± 30 W and ±100W, respectively. For laser power, the smallest change that causes a measurable change in sensing data is 225 W, while for TS, the value is 1.5 mm/sec. The minimum change to cause variation in sensing data is 10 mm/sec for wire feed rate and 100 W for hot wire power.

Figure 6 shows the sensitivity analysis for the change in laser power of ± 500 W and the corresponding change in the molten pool temperature and molten pool size. It can be seen that the change of laser power depicts a considerable variation to be detected by the pyrometer. The step-change of laser power ± 500 W causes a ± 50°C change in the molten pool temperature and a slight increment and decrement of molten size as indicated by Figure 6(b-d). The transient response of molten pool size and temperature to step changes in laser power is also presented. Transient analysis on the WLAM process allows one to determine how long it takes for the process parameters to reach a new steady-state due to step-change to guides process control in real-practice. For example, this information can be used to determine the lead/lag time of control actions in the future. In this case, as seen from Figure 6(b-d), it shows it takes about 3 to 5 seconds to reach the new steady-state of the molten pool after a step-change in laser power.

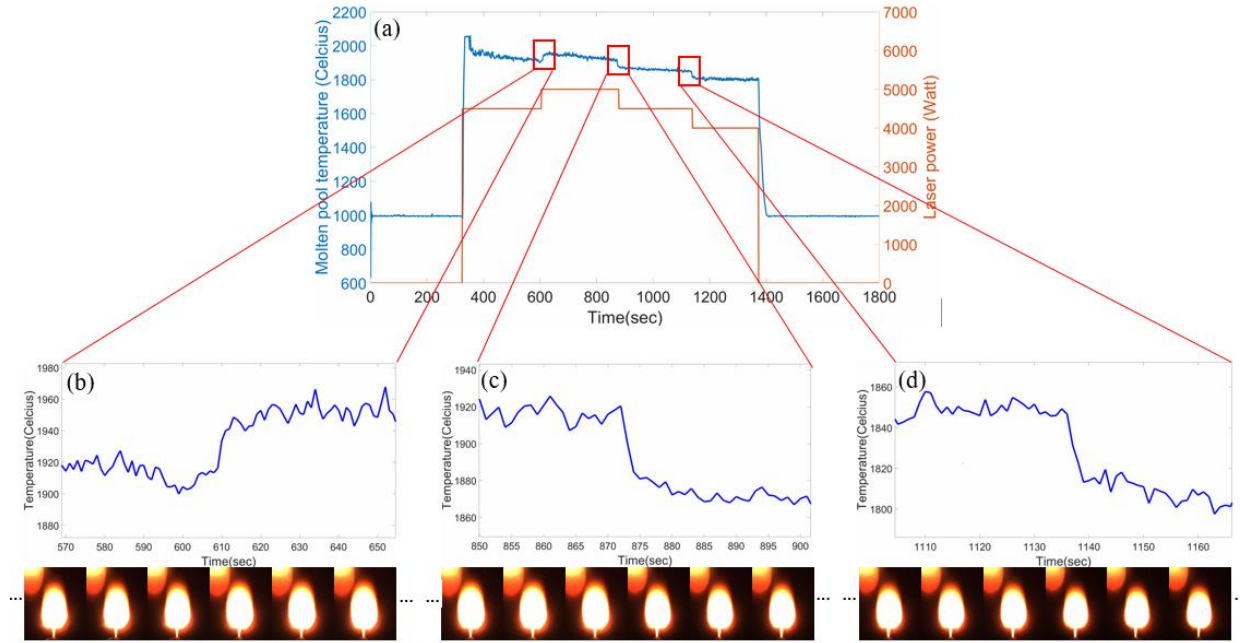

Figure 6. Sensitivity analysis for resolution and prediction analysis (Laser power = ±500W).

*3.3. Feature correlation analysis*

In our case, we have measurable data from four different sensors: one from the camera and three signals from the pyrometers. Table 2 shows the Pearson correlation matrix between the sensing data and process parameters (PP). Analyzing the Pearson correlation between the measurable data and process parameters identified a maximum correlation between the molten pool temperature and the process parameters. Based on this knowledge, the input for model development is the molten pool image and molten pool (MP) temperature. The MP temperature is more correlated with process parameters than the trailing edge (TE) and leading edge (LE) temperature. Hot wire power has the least correlation with both the process parameters and temperature. Therefore, process parameters LP, TS, and WFR are the outputs of the CNN S-P model, while the input to the model is either image data or image data plus MP temperature.

Table 2. Pearson correlation matrix between the sensing data and process parameters.

| **Variable** | **Process parameters** | | | | **Sensing data** | | |
|---|---|---|---|---|---|---|---|
| | **LP** | **TS** | **WFR** | **HWP** | **MP temp** | TE temp | LE temp |
| **LP** | 1 | -0.24 | 0.77 | 3.10E-16 | **0.94** | 0.43 | 0.82 |
| **TS** | -0.24 | 1 | -0.04 | -6.04E-17 | -0.16 | -0.09 | -0.37 |
| **WFR** | 0.77 | -0.04 | 1 | 0.43 | **0.73** | 0.13 | 0.57 |
| HWP | 3.10E-16 | -6.04E-17 | 0.43 | 1 | -0.01 | -0.39 | 0.02 |
| **MP temp** | 0.94 | -0.16 | 0.73 | -0.01 | 1 | 0.40 | 0.76 |
| TE temp | 0.43 | -0.09 | 0.13 | -0.39 | 0.40 | 1 | 0.16 |
| LE temp | 0.82 | -0.37 | 0.57 | 0.02 | 0.76 | 0.16 | 1 |

*3.4. Network convergence*

The single-input multi-output (SIMO) CNN architecture for predicting control parameters using just the image data consists of 31 layers with batch normalization, tanh activation function, and a dropout of 30%. In the present study, the trained architecture for image data as input consists of seven convolution layers interlaced with three global average pooling layers. In contrast, the multi-input multi-output (MIMO), i.e., the multi-modality network, consists of a total of 24 layers, where the first 21 layers are used for image feature extraction. The remaining 3 layers are for predicting the final control parameters along with the temperature data feature. The network uses batch normalization, tanh activation function, and a dropout of 50%. The multi-modality CNN consists of four convolution layers interlaced with two global average pooling layers. Following the second global average pooling, the resulting output is unrolled into a vector and feed into a fully connected layer of dimension 100, which is reduced to a size of 3 before concatenating with the temperature feature. The final layer consists of 3 nodes, based upon the three process parameters' prediction in our case.

A uni-modality and multi-modality CNN architectures are trained using the mean square error as a loss function between the actual and predicted control parameters for the specific build. The evaluation metric for validating the training and testing is the convergence and the accuracy performance of the trained model. The convergence is achieved by calculating the gradient of the loss function and propagating it back through the network. The network weight incoming from the neurons is altered to minimize the loss function during backpropagation. The hyperparameters influencing the backpropagation and controlling the network convergence are the learning rate, momentum, and velocity. The optimization parameters for the uni-modality and multi-modality CNN utilized for training the models are detailed in Table 3.

Table 3. Optimized hyperparameter values for the CNN architectures.

| **Hyper-parameter** | **CNN model** | |
|---|---|---|
| | Uni-modality: Image data | Multi-modality: Image and temperature data |
| Learning rate | $10^{-5}$ | $10^{-5}$ |

| | | | |
|---|---|---|---|
| Momentum | - | | 0.9 |
| Epoch | 10 | | 10 |
| Batch size | 10 | | 10 |

Figure 7 shows the training loss for the CNN architectures during the training stage. The training loss suggests that both the models converge towards the output parameters for the optimized hyperparameters.

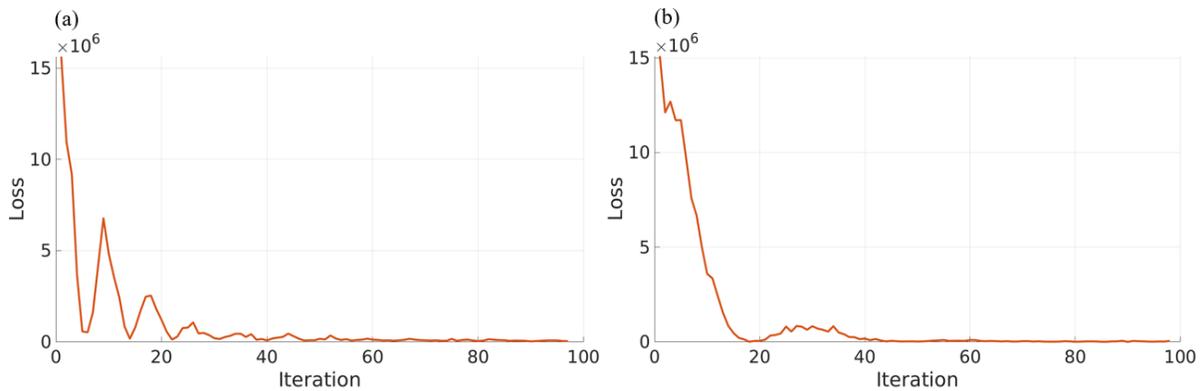

Figure 7. Training loss for uni-modality and multi-modality CNN structure (a) Uni-modality CNN: Image data, (b) Multi-modality CNN: Image and MP temperature data.

## 3.5. Performance prediction

Table 4 shows the results for uni-modality CNN using just the image as input and multi-modality CNN using image and temperature data. The two CNN architectures are trained for an epoch of 10 with 6000 training samples and 500 test samples of image and temperature data collected during 13 builds. The CNN architecture uses 6-fold cross-validation and the results discussed below are for the unseen test dataset.

Table 4. Comparison between uni-modality and multi-modality CNN for predicting the control parameters using RMSE, RE, NRMSE, and SD.

| Process parameter | CNN model | | | | | | | |
|---|---|---|---|---|---|---|---|---|
| | Uni-modality: Image data | | | | Multi-modality: Image and MP temperature data | | | |
| | RMSE | RE | NRMSE | SD | RMSE | RE | NRMSE | SD |
| LP | 385.23 | 7.48% | 19.26% | 192.08 | 186.48 | 3.49% | 9.32% | 115.58 |
| TS | 1.54 | 37.70% | 23.77% | 1.15 | 1.12 | 26.70% | 17.29% | 0.77 |
| WFR | 7.97 | 16.95% | 25.47% | 3.15 | 6.41 | 13.93% | 20.51% | 3.92 |

The network performance is compared using RMSE, RE, and NRMSE for the three predicted control parameters. As seen from Table 4, the RMSE values for the multi-modality CNN are below the sensitivity values for the corresponding process parameters allowing the model's output prediction to be used in control architecture.

Table 4 depicts the fact that the relative error for travel speed is high compared to laser power and wire feed rate for both uni-modality and multi-modality CNN. The high percentage value is caused by the lower end of travel speed, i.e., 3.5 mm/s, which is significantly lower than the normalized travel speed of 6.6 mm/s. For a laser DED process, laser power and travel speed are highly correlated parameters based on analytical and numerical modeling. However, feature selection using the Pearson correlation for the current dataset shows no such relation because of data sparsity. The upper and lower bound for the TS, i.e., 3.5-10 mm/s, is smaller, and there exists an improper distribution of travel speed within the set bound. Hence, the high RE is depicted for the travel speed.

Figure 8 shows the RMSE and SD comparison for the developed uni-modality and multi-modality CNN architecture. It can be seen from Table 4 and Figure 8 that the standard deviation for the multi-modality CNN is lower compared to the uni-modality CNN in the case of LP and TS. The SD for LP is 192.08 and 155.58 for uni-modality and multi-modality CNN, respectively. TS has a standard deviation value of 1.15 for the uni-modality network and 0.77 for the multi-modality CNN. The SD for WFR is 3.15 and 3.92 for uni-modality and multi-modality CNN, respectively. Although the SD for the WFR is higher for multi-modality CNN, the RMSE is lower.

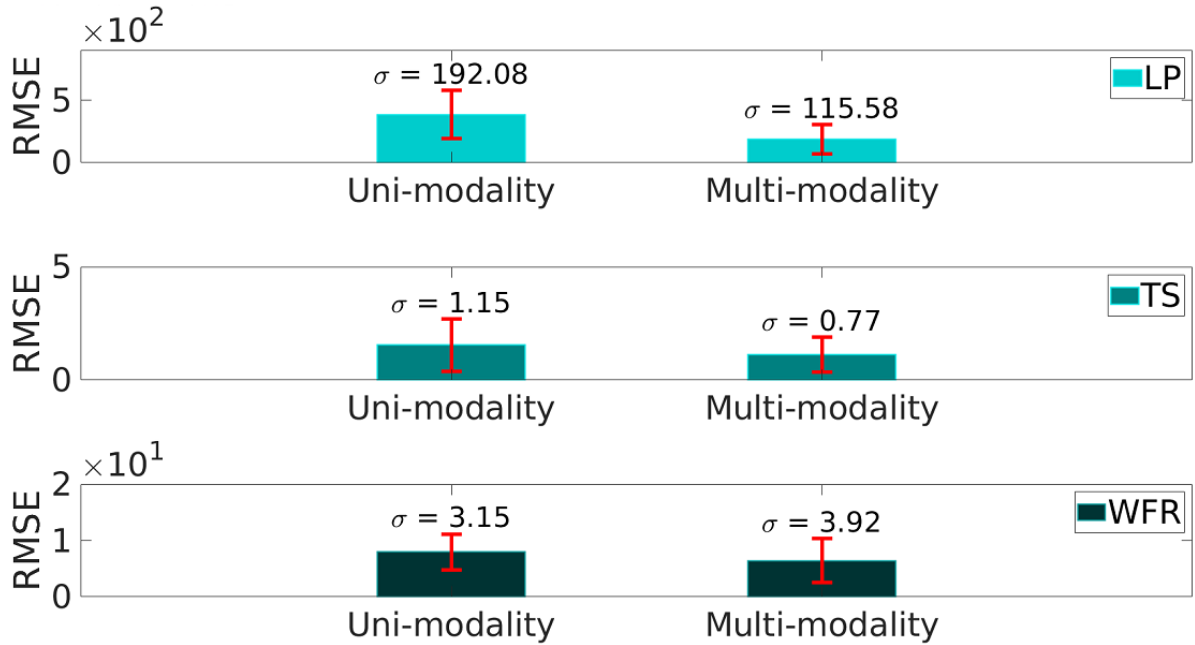

Figure 8. RMSE and SD comparison for uni-modality and multi-modality CNN.

Current work shows a CNN approach that predicts the current control parameters required for the desired sensor data, i.e., molten pool image and temperature. The results have shown that adding molten pool temperature as a feature to the image data improves the model's performance by 51.6%, 27.3%, and 19.5% for laser power, travel speed, and wire feed rate, respectively. Data sparsity in the current dataset cannot accurately capture the underlying physics for the molten pool condition-process relations model using single-modality.

## 4. Discussions

### 4.1. Molten pool variations with the change in process parameters

The WLAM process on its own is 'open-loop' with an end-user providing pre-defined process parameters. As mentioned, these process parameters are often tailored via trial-and-error for a specific machine and material combination, and later modifications are made to the process parameters manually based on measured outcomes. It has been shown that time-invariant process parameters result in parts with an anisotropic microstructure, which can lead to variations in properties within the fabricated part, and post-heat treatments can sometimes eliminate variability but not always [56]. The utilization of constant process parameters can lead to thermal distortion, increased residual stress, and cracking [4, 14]. For example, residual heat from previous layers can make subsequent layers have a larger and larger molten pool and/or higher molten pool temperature when the laser power is held constant. An element of feedback control is needed to be integrated into the WLAM system by monitoring or detecting changes to molten pool morphology and temperature. Understanding the process parameters and their impact on the molten pool evolution is therefore of primary importance. The WLAM system can adjust the process parameters or control the idle time between layers in a timely fashion as the process progresses.

As seen from Figures (3-5), in the steady-state regime, we experimentally observed the effects of process parameters on the molten pool thermal and morphology characteristics. The detected signals show dependence on the main process parameters, including laser power, travel speed, and wire feed rate. The sensors' signals are suitable for monitoring and building a future closed-loop control framework in the developed WLAM system. To efficiently and effectively control the molten pool and process variables, it is also essential to understand the molten pool's transient response to step changes in laser power or travel speed [4]. Parametric sensitivity shows how the input uncertainty affects the output response and its utilization for system modeling and control decisions. Sensitivity analysis in section 3.2 is performed to quantify the influence of step changes in the process parameters on the molten pool's behavior.

*4.2. Molten pool temperature influence on model's performance*

Raghavan et al. [44] developed a heat transfer and fluid flow model for the laser DED process to examine how changes in processing parameters affect the relationships between molten pool surface, thermal cycles, and solidification parameters in Ti-6Al-4V. Molten pool surface area is found to be a poor indicator of cooling rate at different locations in the molten pool. Therefore, it cannot be relied upon to achieve targeted part quality control. It is shown that monitoring or controlling only the molten pool surface area can be insufficient for achieving targeted microstructures and mechanical properties. As indicated in Table 4, the multi-modality CNN model combining additional temperature features outperforms the uni-modality CNN model, providing more comprehensive control of the WLAM process. Performance improvement is mainly because monitoring the surface molten pool geometry does not provide sufficient information to incorporate molten pool depth and temperature information. The geometric accuracy of the printed part depends on the coupling effects of bead height, width, and depth of the molten pool. The printed part's mechanical property and defect population are sensitive to the molten pool's thermal history.

Based on the Pearson correlation matrix from Table 2, the molten pool temperature is more likely to correlate with process parameters compared to the leading edge and trailing edge temperature of the solid material just in front of and behind the molten pool. Table 5 further verifies that the multi-modality model's prediction performance using molten pool image plus MP temperature generally outperforms the other two models in terms of error metrics, i.e., RMSE and RE. There is a location dependent temperature profile and gradient across the fusion depth and radius of the molten pool. A near-gaussian temperature profile exists in the molten pool with the highest value near the laser location and decreasing linear variation toward the molten pool's trailing/leading edge [57, 58]. From our experiments at various process parameters, almost all of the molten pool's temperature at the center of the molten pool is greater than 1800˚C (Figures (3-5)), which is overall greater than the melting point of Ti-6Al-4V being about 1660˚C. In comparison, the trailing/leading edge temperature is less than the molten pool temperature in the range of 250-400˚C. The temperature of the molten pool at the center near the laser heating location is a more representative temperature feature for the molten pool during printing, providing more useful information for sensors-process modeling.

Table 5. Comparison between different temperature input to the multi-modality CNN for predicting the control parameters using RMSE, RE, NRMSE, and SD.

| PP | CNN model | | | | | | | | | | | |
|---|---|---|---|---|---|---|---|---|---|---|---|---|
| | Multi-modality: Image and LE temperature data | | | | Multi-modality: Image and MP temperature data | | | | Multi-modality: Image and TE temperature data | | | |
| | RMSE | RE | NRMSE | SD | RMSE | RE | NRMSE | SD | RMSE | RE | NRMSE | SD |
| LP | 472.28 | 9.05% | 23.61% | 340.13 | 186.48 | 3.49% | 9.32% | 115.58 | 632.69 | 11.56% | 31.63% | 336.57 |
| TS | 1.13 | 24.94% | 17.37% | 0.70 | 1.12 | 26.70% | 17.29% | 0.77 | 1.14 | 26.69% | 17.52% | 0.73 |
| WFR | 10.76 | 22.59% | 34.37% | 4.42 | 6.41 | 13.93% | 20.51% | 3.92 | 7.55 | 16.11% | 24.12% | 3.67 |

Table 5 shows the multi-modality CNN comparison for different temperature measurements as input to the network. The leading edge, molten pool, and trailing edge temperature are used as input to the network along with the image data. The three CNN architectures are compared using RMSE, RE, NRME, and SD.

*4.3. S-P model for in situ quality assurance*

The trained S-P model can be used as a feedforward controller for controlling the molten pool geometry and temperature. Maintaining a consistent value of molten pool geometry and temperature allows for potentially achieving desired final build geometry and microstructural properties needed for quality assurance. Figure 9 represents the in situ quality assurance framework based upon the developed sensing-process (S-P) model. The molten pool sensing data will be fed into the trained S-P model to predict the corresponding process parameters, which in turn can provide the desired build quality. The data-driven S-P model serves as a feedforward controller, providing predictive process parameter input to the system to achieve the desired molten pool geometry and temperature.

The machine setting for process parameters often deviates from the default set value or has fluctuations. A proportional integral derivative (PID) feedback controller can be employed to deal with unmeasured disturbances caused by system operation. The feedback functions by continuously measuring the system's output and providing corrective input calculated from the PID control algorithm. For in situ molten pool geometry and temperature control, the process parameter predictions from the S-P model can be used in conjunction with the feedback controller for computing the ultimate process parameters to be used by the printer. In this way, the designed architecture can serve as an in situ quality assurance control framework for the laser DED system.

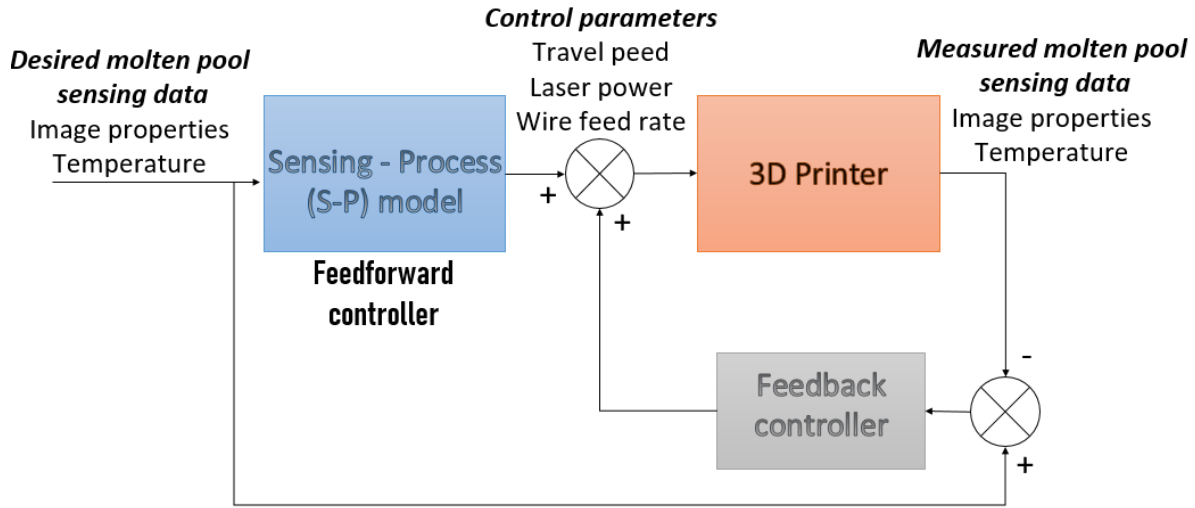

Figure 9.  Proposed architecture for a laser wire feed DED process based on machine learning using the S-P model for in situ quality assurance.

## 5. Conclusion

In this paper, we collected sensor data from the molten pool in a WLAM process under a set of controlled input process parameters. A vision and a pyrometer-based sensing system are instrumented for laser hot wire feed direct energy deposition to acquire molten pool sensing data. The build data set consists of different process parameters for a single bead deposition. The molten pool-process relation is described and analyzed experimentally along with the Pearson correlation matrix.  A molten pool-process parameter model is developed using CNN to map the relationship between the molten pool sensing data and process parameters as a part of the in situ quality assurance framework. The multi-modality CNN is used to predict the control parameters using only the image data or both image and temperature sensors' data as input. Temperature is added as an external feature to the features extracted from the image data for process parameter prediction. The network is optimized for the structure, including the input and output using feature selection. The prediction results indicate that adding temperature for process parameter prediction improved the model's performance by adding a different modality to the network. The multi-modality model shows an improvement of 51.6% in laser power, 27.3% in travel speed, and 19.5% in wire feed rate. The results reveal that CNN is a useful technique to build an end-to-end prediction model directly from the measurable sensor data with minimal preprocessing. The trained S-P model can be used as a feedforward controller for providing predictive actions to the system based upon experimental data knowledge.

**Conflict of interest**
The authors do not have any conflict of interest.